\documentclass[preprint,nofootinbib,eqsecnum,pre,12pt]{revtex4}
\usepackage{graphicx}
\usepackage{dcolumn}
\usepackage{bm}
\usepackage{amsfonts}
\usepackage{amsmath}
\usepackage{amssymb}
\usepackage{graphicx}
\newcommand{\eref}[1]{Eq.~(\ref{#1})}
\begin{document}

\title{Comparison of  three-dimensional and  two-dimensional statistical \\
mechanics of shear layers for flow between two parallel plates.}
\author{L. Shirkov$^{\dagger}$, V. Berdichevsky$^{\ddagger}$}
\affiliation{$^{\dagger}$Department of Chemistry, University of Warsaw, Poland\\
$^{\ddagger}$Mechanical Engineering, Wayne State University, Detroit MI 48202}

\date{Received: date / Accepted: date}

\begin{abstract}
It is shown that the averaged velocity profiles predicted by statistical
mechanics of point vortices and statistical mechanics of vortex lines are
practically indistinguishable for a shear flow between two parallel walls.
\end{abstract}

\maketitle

\section{\label{sec:level1}Introduction}

It is known that statistical mechanics of point vortices describes
surprisingly well the averaged velocity profiles of self-similar mixing
layer. After development of statistical mechanics of vortex lines \cite
{rf1,rf2,rf3,rf4,rf5}, where a vortex does not remain straight and is
allowed to take wavy shapes in the course of motion, it appeared a concern
that the above-mentioned feature of statistical mechanics of point vortices
can be lost in three-dimensional theory. In this paper we show that this is
not the case: the averaged velocity profiles predicted by statistical
mechanics of point vortices and statistical mechanics of vortex lines are
practically indistinguishable for a shear flow between two parallel walls.

\section{Averaged equations for flow between two plates}

Consider a flow of ideal incompressible fluid between two parallel walls, $y$
is the coordinate normal to the walls, $-h\leq y\leq h$, $2h$ - the distance
between the walls. The flow is modeled by motion of a large number of
vortices (Fig.1). Flow is periodic in $x$-direction and $z$-direction. In
the limit of infinite period in $z$-direction, the averaged velocity is
parallel to the walls, does not depend on $x$ and has the only non-zero
component, $u=u(y)$. Assuming that vortices have the same intensity and the
total discharge is zero, one obtains for the stream function of the averaged
flow, $\psi=\psi(y)$ ($u\equiv d\psi/dy$), the equation

\begin{equation}
\frac{d^{2}}{dy^{2}}\psi=-\sigma f(y),\quad\left. \frac{d}{dy}%
\psi\right\vert _{\pm h}=\pm U,
\label{eq:general}
\end{equation}
where $f(y)$ is the probability to find a vortex at the point $y$, $\sigma$
is the total vorticity per unit length in $x$-direction, $\sigma=-2U$. In
the case of point vortices (Fig.1a), \cite{rf6,rf7} 
\begin{equation}
f\left( y\right) =\frac{e^{-\beta\sigma\psi\left( y\right) }}{\int
_{-h}^{h}e^{-\beta\sigma\psi\left( y^{\prime}\right) }d^{2}y^{\prime}},
\label{eq:prob1}
\end{equation}
and equations \eref{eq:general}, \eref{eq:prob1} form a closed system of equations. This system can be
solved analytically. Parameter $\beta$ has the meaning of inverse
temperature of vortex motion. It is determined by the initial energy of
turbulent flow.\newline
In the case of deforming vortex lines (Fig.1b), $f(y)$ is expressed through
the solutions of the eigenvalue problem, \cite{rf4,rf5}

\begin{figure*}[tbp]
\includegraphics[width=0.8\textwidth]{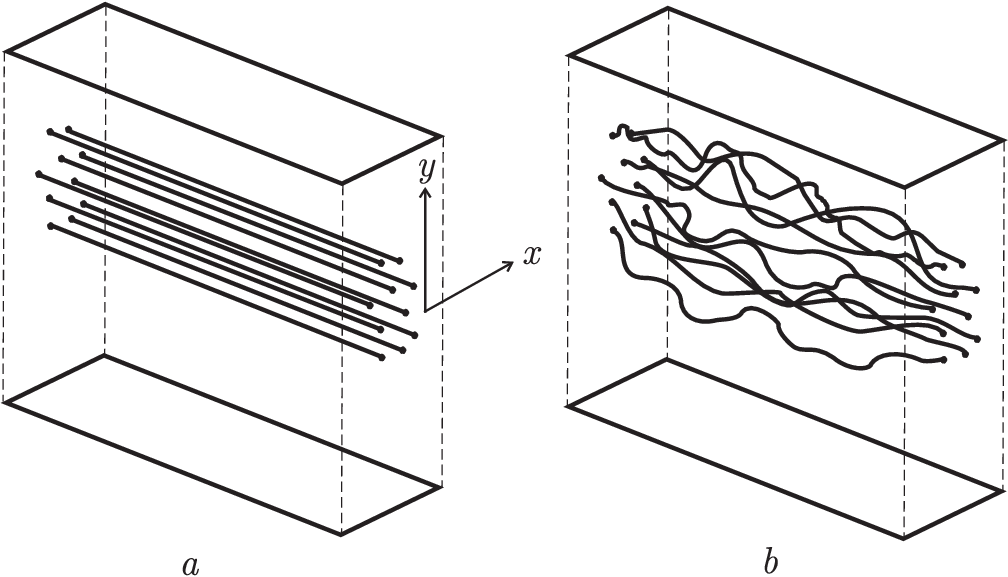} 
\caption{Two-dimensional flow of point vortices ($a$) and three-dimensional
flow of vortex lines ($b$)}
\label{fig:1}
\end{figure*}

\begin{equation}
\Delta\varphi-\beta\sigma\psi\varphi=-\lambda\varphi,~~~\quad\left. \frac {d%
}{dy}\varphi\right\vert _{\pm h}=0,
\label{eq:ev}
\end{equation}
$\lambda$ being the minimum eigenvalue. Similarly to quantum mechanics, $%
f(y) $ is proportional to the squared solution of the eigenvalue problem: 
\begin{equation}
f(y)=\frac{\varphi^{2}(y)}{\int_{-h}^{h}\varphi^{2}(y)dy^{\prime}}.
\label{eq:prob2}
\end{equation}
We aim to compare solutions of \eref{eq:general},\eref{eq:prob1} and \eref{eq:general}, \eref{eq:ev}, \eref{eq:prob2}.

\begin{figure*}[tbp]
\centering
\includegraphics[width=0.34\textwidth,angle=270]{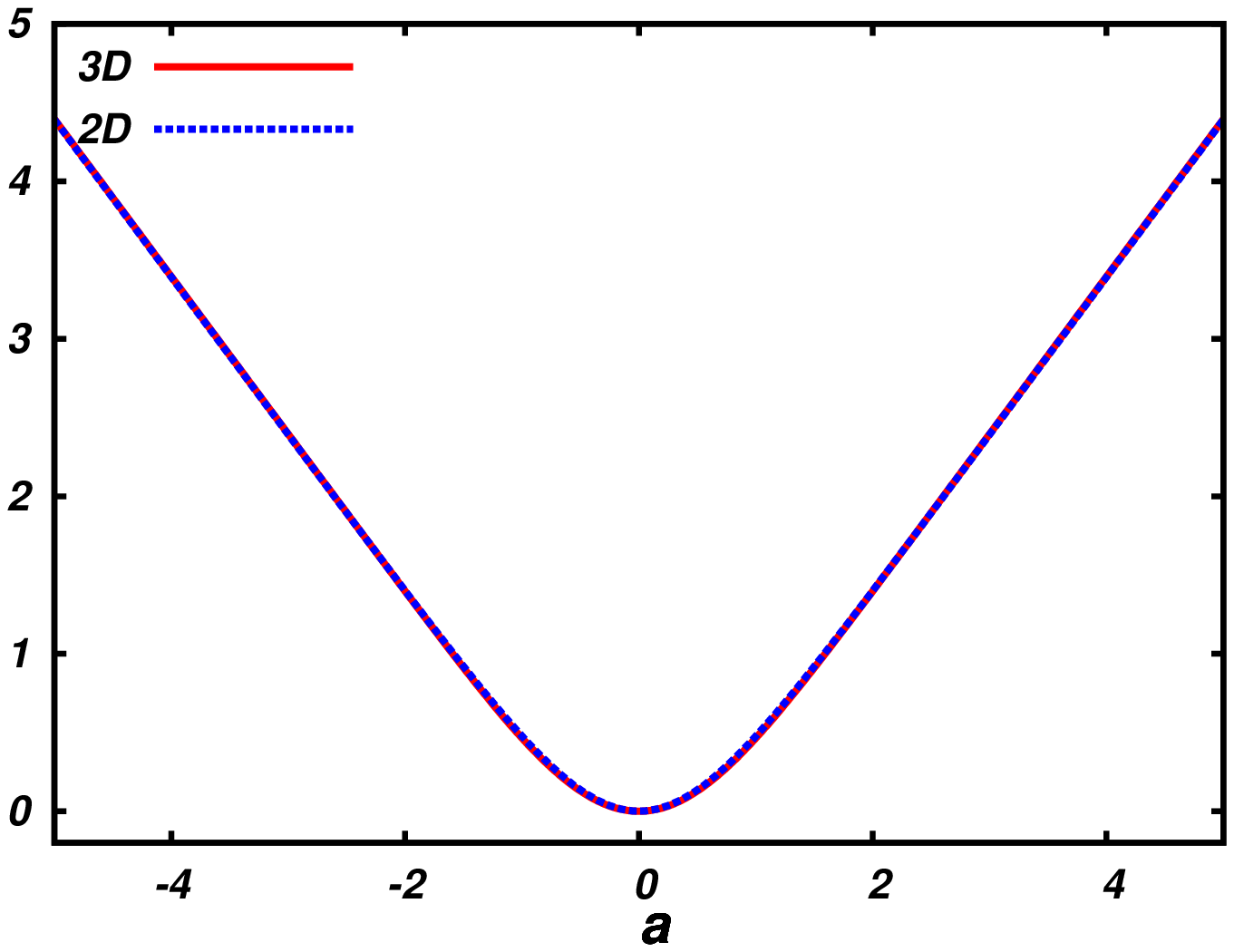} %
\includegraphics[width=0.34\textwidth,angle=270]{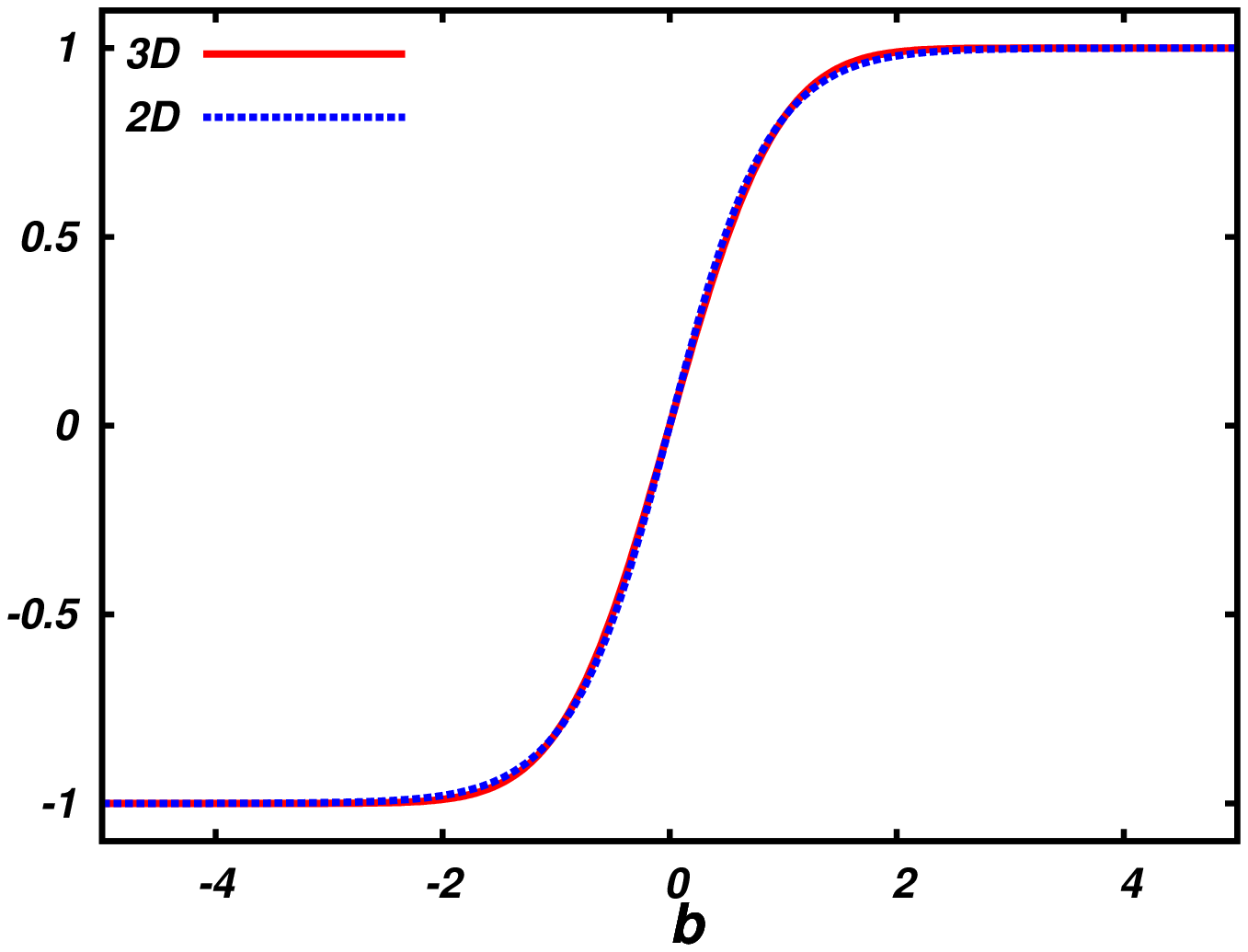} 
\caption{Stream function ($a$) and velocity ($b$) profiles for $\protect\beta%
_{3D}=-1.0$ and $\protect\beta_{2D}=-1.1374$ }
\label{fig:2}
\end{figure*}

\section{Results and discussion}

\label{sec:1}

There is no reason to expect that the velocity profiles found from the two
quite different system of equations, \eref{eq:general}-\eref{eq:prob1} and \eref{eq:general},\eref{eq:ev},
\eref{eq:prob2}, coincide.
Nevertheless, this turns out to be the case: the velocity profiles are
practically indistinguishable. More precisely: for each $\beta$ from "$3D$
problem" \eref{eq:general},\eref{eq:ev},
\eref{eq:prob2}, there is $\beta$ from "$2D$ problem" \eref{eq:general}-\eref{eq:prob1} for which
the velocity profiles practically coincide.

System of equations \eref{eq:general}-\eref{eq:prob1} admits an analytical solution: 
\begin{equation}
\psi =-\frac{1}{U\beta }\ln \cosh (Cy),
\label{eq:sol2d_1}
\end{equation}%
where the constant $C$ is determined from the boundary condition, 
\begin{equation}
U+\frac{C}{U\beta }\tanh (Ch)=0.
\label{eq:const}
\end{equation}%
For $h=\infty $, $C=-U^{2}\beta $ and \eref{eq:sol2d_1} becomes: 
\begin{equation}
\psi =-\frac{1}{U\beta }\ln \cosh (U^{2}\beta y).
\label{eq:sol2d_2}
\end{equation}

All parameters of the flow can be normalized with respect to the wall
velocity: $\psi\rightarrow\psi/U$, $\sigma\rightarrow\sigma/U$ and $%
\beta\rightarrow\beta U^{2}$. The dimension of the normalized $\beta$ is
length$^{-1}$. The characteristic length, $\left\vert \beta\right\vert
^{-1}, $ is the width of the mixing layer.

Non-linear eigenvalue problem \eref{eq:general},\eref{eq:ev},
\eref{eq:prob2} was solved numerically using an
iteration procedure. At each iteration step the system is being reduced to
the Sturm-Liouville problem, that was solved with the Pr\"{u}fer method (a
shooting method based on oscillation \cite{rf8,rf9}). For a given $\beta $
of vortex line flow, $\beta _{3D},$ one can choose $\beta $ of point vortex
flow, $\beta _{2D},$ in such a way, that the velocity profiles are
practically identical. This can be seen, for example, from Fig.2, where the
velocity profiles and the stream function are shown for $\beta _{3D}=-1.0000$
and $\beta _{2D}=-1.1374$. The values of $\beta _{2D}$ and $\beta _{3D}$ for
which the average velocity profiles coincide form a curve in the plane \{$%
\beta _{2D},\beta _{3D}\}$ shown in Fig.3. This curve was found in the
following way. For each $\beta _{3D}$ we seek $\beta _{2D}$ by the
minimization of the sum%
\begin{equation}
\sum_{i=1}^{N}w_{i}|u_{i}^{2D}-u_{i}^{3D}|,
\label{eq:sum}
\end{equation}%
%
%
where $u_{2D}$ and $u_{3D}$ are velocities in $2D$ and $3D$ problems,
respectively, $N$ is the number of the mesh points. The weight coefficients, 
$w_{i},$ depend on the density of the mesh and the gradient of the velocity
profile $u$. The distance between the walls, $h,$ was chosen large enough
for the solution \label{eq:sol2d_2} to be applicable.

\begin{figure}[tbp]
\centering
\includegraphics[width=0.40\textwidth,angle=270]{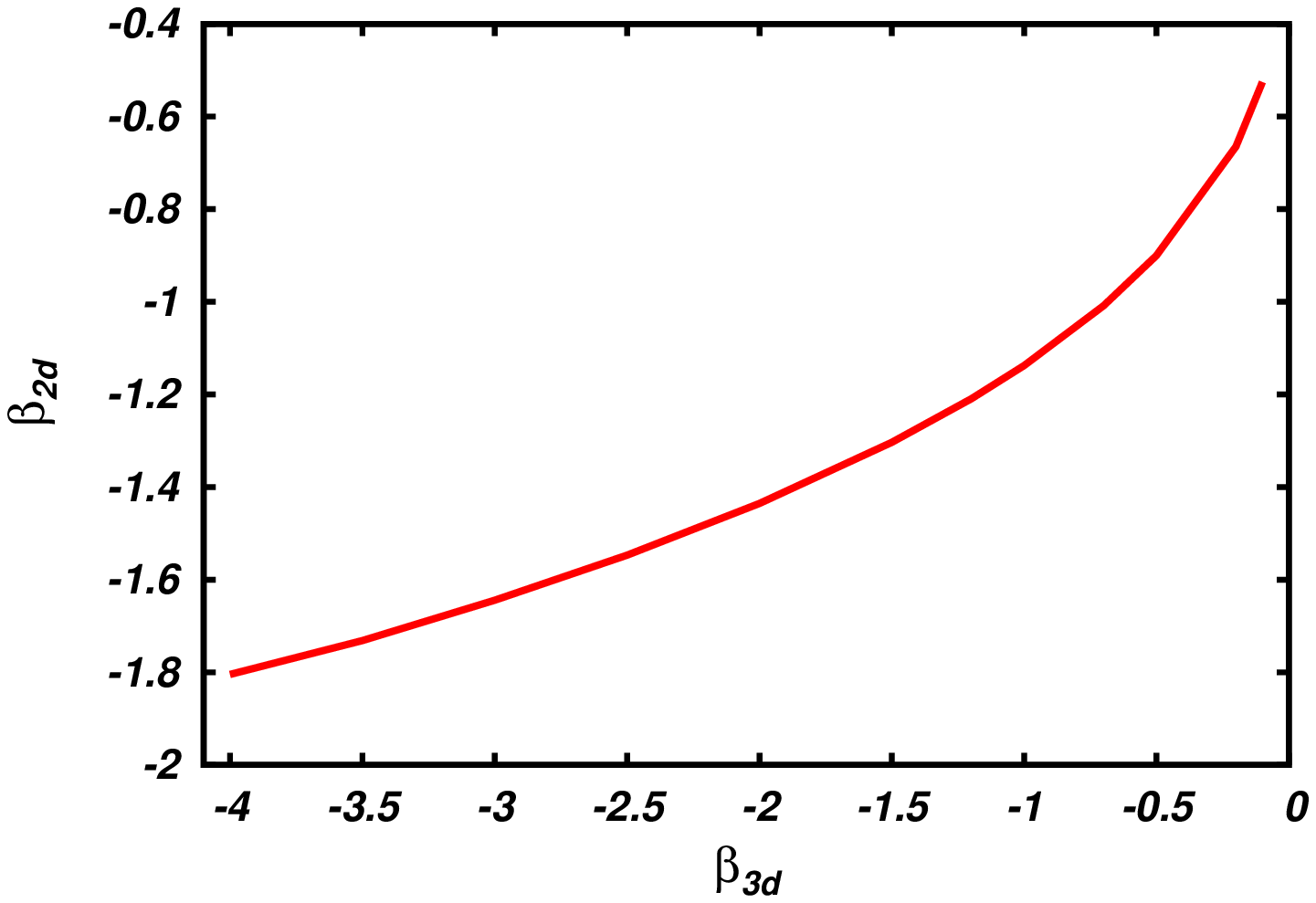} 
\caption{Two-dimensional inverse temperature $\protect\beta_{2D}$ vs.
three-dimensional inverse temperature $\protect\beta_{3D}$ }
\label{fig:3}
\end{figure}

The parameters $\beta _{2D}$ and $\beta _{3D}$ have the meaning of the
inverse temperature of two-dimensional and three-dimensional motions. The
fact, that the same velocity profile corresponds to different $\beta _{2D}$
and $\beta _{3D}$, indicates that the corresponding temperatures of
two-dimensional and three-dimensional motions are different. The temperature
of two-dimensional motion has a simple physical meaning: this is an average
area bounded by the vortex trajectory. The area has orientation.
Accordingly, temperature may have both signs. Negative temperature of the
flow considered corresponds to clockwise pass of the curls of the vortex
trajectories \cite{rf1}. The physical meaning of temperature of
three-dimensional vortex line motion is not known, therefore a physical
interpretation of the graph $\beta _{2D}(\beta _{3D})$ is yet to be
established. The result obtained seems an indication that for the shear flow
between parallel walls three-dimensionality does not play an important role
in formation of the averaged velocity profiles.

This paper has been published previously in the Russian journal \cite{rf9} that is not distributed in the West and on internet. 


\end{document}